\documentclass[prl,twocolumn,showpacs,preprintnumbers]{revtex4}
\usepackage{graphicx}
\usepackage{amsmath}
\usepackage{amssymb}
\usepackage{times}

\begin{document}

\noindent{\bf Sakai, Motome, and Imada Reply:}
In Ref.~\cite{sm09}, we have clarified the structure of
poles and zeros of Green's function for the 2D Hubbard model
in the full momentum-energy space by using a cluster
extension of the dynamical mean-field theory, and
have elucidated that the interference between 
poles and zeros
plays a crucial role in the doped Mott insulator.
The study provides a comprehensive understanding of
a number of unconventional features of Mott physics
such as hole pockets, pseudogap, Lifshitz transitions,
non-Fermi liquids, Fermi arcs, and 
a spectral weight transfer over the Mott gap. 

In the preceding Comment \cite{p09} on our Letter \cite{sm09}, 
Phillips claims about only one issue confined to
the spectral weight transfer, among our wide-ranging clarifications. 
The criticism is that we consider an excess spectral weight 
just above the chemical potential in doped states 
as a non-trivial finding
even though it has already been discussed in the prior works \cite{c91,em91,me93}. 
Phillips also criticizes that our argument referring to the double occupancy $n_d$ is not compatible with the perturbation theory \cite{hl67} and offers no underlying physical ground.

The former comment has nothing to do with our contribution 
from a careful analysis of updated numerical results. 
As clearly stated in our Letter \cite{sm09},
our emphasis is not on the excess weight itself 
but on its sharp rise by tiny doping. 
As shown in Fig.~3(b) of Ref.~\cite{sm09}, our numerical data indicate
that the spectral weight $W_1$ ($\Lambda$ in Ref.~\cite{p09})
increases quite rapidly as a function of doping $\delta$
in the vicinity of the Mott transition.
Phillips also criticized that our result is identical to that in Fig.~2 of Ref.~\cite{me93}, 
but this is totally incorrect: The result in Ref.~\cite{me93} is for 1D system,
and furthermore, such rapid increase near $\delta=0$ is not observed
even though there is an additional weight to the static component proportional to $\delta$.
The very rapid, non-linear behavior found in our result is quite unexpected, and at least, 
to our best knowledge, has not been anticipated in previous studies.

As to the latter criticism on the incompatibility of ours with the perturbation result
with respect to $t/U$, 
it is indeed difficult to expect that such abrupt change in $W_1$
is described by perturbative arguments. 
Although Phillips criticized the incompatibility, 
the first-order perturbation, 
which is reproduced in Eq.~(1) in Ref.~\cite{p09},
 trivially fails as $\delta \to 0$,
because the kinetic energy density approaches a nonzero value for finite $U$.
The incompatibility should indeed be there, 
as Phillips also realized in his next paragraph. 
Precisely, the quick rise of $W_1$ from zero is the point which 
delivers the significance of the non-trivial finding. 

Beyond the quick rise, $W_1$ appears to be consistent with a scaling 
$\delta +n_d$
which we proposed by considering the change in double occupancy $n_d$. 
The overall behavior of $W_1$ in this range of $\delta$ can be 
reproduced by either our scaling $W_1 \simeq \delta + n_d$ 
or the perturbation result \cite{hl67}. 
This is reasonably understood by noting the fact that, as also mentioned in Ref.~\cite{pc09},
the kinetic energy term in the perturbative expansion is closely related 
to the doublon density $n_d$
because the former creates doublon-holon pairs in the atomic limit.

To explain the non-linear evolution of $W_1$ at tiny doping followed by
the scaling $\delta +n_d$ at higher doping, we need a new mechanism.
It motivated us, in Ref.~\cite{sm09}, to propose
an avalanchine screening for doublon-holon bound pairs.
Once holes are doped into the Mott insulator whose gap is determined
by the doublon-holon binding,
the mobile holes bring about the screening of the binding energy.
The screened binding energy survives as the pseudogap. 
What we proposed is that the screening should have a positive feedback 
accelerating the weight transfer,
since dissolved bound pairs further join in the screening process.
This feedback will be particularly effective in 2D, 
because the screening is governed 
by the density of states at the Fermi level, which jumps to a nonzero value 
in 2D by doping with the formation of Fermi surface pocket. 
This self-accelerated process
brings about the abrupt reduction of the doublon-holon binding
and the non-linear increase of $W_1$ upon doping.
In Ref.~\cite{sm09}, it was inferred to drive the system to the verge of a first-order transition or a marginal quantum criticality.

In Comment \cite{p09}, Phillips claimed that we do not offer any 
explanation for the underlying mechanism 
of the quick reduction of the Mott gap associated with $W_1$.
On the contrary, as we repeated above, we already gave a picture in Ref.~\cite{sm09} which is qualitative but physically sound and, at least,
reasonably fits to our numerical results.

We reached, in Ref.~\cite{sm09}, the conclusion that
interplay between doublon-holon bound pairs and mobile carriers
is a key for understanding the Mott physics,
as also addressed within a scenario of ``2e-boson" recently hypothesized 
in Ref.~\cite{pc09} and Comment~\cite{p09}. 
To clarify the binding mechanism with quantum fluctuations and its variation as doping
will give deeper understanding of the Mott physics
beyond the prior perturbative arguments in the atomic limit.
It would also be an intriguing issue how our 
wide-ranging results including the hole-pockets, 
Fermi arcs, Lifshitz transitions and other nontrivial findings in Ref.~\cite{sm09}
obtained by the state-of-the-art numerical calculations 
can be reproduced in the ``2e-boson" calculations in Ref.~\cite{pc09}.   

\medskip

\noindent Shiro Sakai, Yukitoshi Motome, and Masatoshi Imada

\small{\noindent Department of Applied Physics\\
       University of Tokyo\\
       Hongo, Tokyo 113-8656, Japan}


\begin{references}
\item[\nonumber]
\bibitem{sm09}
S. Sakai, Y. Motome, and M. Imada, Phys. Rev. Lett. {\bf 102}, 056404 (2009).
\bibitem{p09}
P. Phillips, arXiv:0904.0454.
\bibitem{c91}
C. T. Chen, {\it et al.}, Phys. Rev. Lett. {\bf 66}, 104 (1991).
\bibitem{em91}
H. Eskes, M. B. J. Meinders, and G. A. Sawatzky, Phys. Rev. Lett.
{\bf 67}, 1035 (1991)
\bibitem{me93} 
M. B. J. Meinders, H. Eskes, and G. A. Sawatzky,
Phys. Rev. B {\bf 48}, 3916 (1993). 
\bibitem{hl67}
A. B. Harris and R. V. Lange, Phys. Rev. {\bf 157}, 295 (1967).
\bibitem{pc09}
P. Phillips, T.-P. Choy, and R. G. Leigh, Rep. Prog. Phys. 
{\bf 72}, 036501 (2009).
\end{references}
\end{document}